\documentclass[journal]{IEEEtran}
\usepackage{amsmath,amsfonts,amssymb}
\usepackage{algorithmic}
\usepackage{algorithm}
\usepackage{array}
\usepackage[caption=false,font=footnotesize,labelfont=sf,textfont=sf]{subfig}
\usepackage{textcomp}
\usepackage{stfloats}
\usepackage{url}
\usepackage{verbatim}
\usepackage{graphicx}
\usepackage{cite}
\usepackage{xcolor}
\usepackage{hyperref}
\hypersetup{
    colorlinks=true,      
    linkcolor=blue,      
    citecolor=blue,
    urlcolor=blue        
}

\begin{document}

\title{Subband Architecture Aided Selective \\Fixed-Filter Active Noise Control}

\author{Hong-Cheng Liang, Man-Wai Mak, \IEEEmembership{Senior Member, IEEE}, and Kong Aik Lee, \IEEEmembership{Senior Member, IEEE}
\thanks{This work was supported in part by Innovation Technology Co., Ltd., under project N-ZDF9.}
\thanks{The authors are with the Department of Electrical and Electronic Engineering, The Hong Kong Polytechnic University, Hong Kong (e-mail: honliang@polyu.edu.hk; enmwmak@poly.edu.hk; kong-aik.lee@polyu.edu.hk).}
\thanks{The code link of our proposal is available at \href{https://github.com/Amao-Liang/Subband-Architecture-for-Selective-Fixed-Filter-Active-Noise-Control}{https://github.com/Amao-Liang/Subband-Architecture-for-Selective-Fixed-Filter-Active-Noise-Control}.}}

\markboth{Preprint}
{Subband Architecture Aided Selective Fixed-Filter Active Noise Control}

\maketitle

\begin{abstract}
The feedforward selective fixed-filter method selects the most suitable pre-trained control filter based on the spectral features of the detected reference signal, effectively avoiding slow convergence in conventional adaptive algorithms. However, it can only handle limited types of noises, and the performance degrades when the input noise exhibits non-uniform power spectral density. To address these limitations, this paper devises a novel selective fixed-filter scheme based on a delayless subband structure. In the off-line training stage, subband control filters are pre-trained for different frequency ranges and stored in a dedicated sub-filter database. During the on-line control stage, the incoming noise is decomposed using a polyphase FFT filter bank, and a frequency-band-matching mechanism assigns each subband signal the most appropriate control filter. Subsequently, a weight stacking technique is employed to combine all subband weights into a fullband filter, enabling real-time noise suppression. Experimental results demonstrate that the proposed scheme provides fast convergence, effective noise reduction, and strong robustness in handling more complicated noisy environments.
\end{abstract}


\section{Introduction}
\IEEEPARstart{A}{coustic} noise increasingly disrupts modern infrastructure and daily life, making the development of effective noise suppression techniques a critical research topic \cite{ref1,ref2}. Distinct from passive noise control \cite{ref3}, active noise control (ANC) plays an irreplaceable role in attenuating low- and mid-frequency noise, with a lot of applications in consumer headsets, aircraft cabins, and electric vehicles \cite{ref4,ref5,ref6}.  

To implement an ANC system for industrial use, adaptive filters \cite{ref7} are widely employed due to their effectiveness. Among them, the filtered-x least-mean-square (FxLMS) and filtered-x normalized least-mean-square (FxNLMS) algorithms are commonly adopted to compensate for the secondary path \cite{ref8}. However, adaptive finite-impulse-response (FIR) filters with hundreds of taps suffer from intensive computation and slow convergence. To overcome these issues, subband adaptive filtering (SAF) techniques \cite{ref9,ref10,ref11,ref12,ref13,ref14,ref15,ref16} have been introduced into ANC systems. By decomposing the input and error signals into multiple bands, both the filter length and update rate can be decimated in each subband, thereby lowering the computational load and accelerating convergence.  

Recently, many solutions exploit control filters with fixed coefficients \cite{ref17,ref18,ref19,ref20}, which can further reduce the complexity requirement and eliminate the risk of divergence. Selective fixed-filter active noise control (SFANC) was proposed in \cite{ref17}. In this method, a series of training noises is first input into the system to obtain a collection of optimal control filters. The coefficients of these filters, along with the spectral characteristics of the corresponding training noises, are stored in a database. During the control stage, the spectral information of the incoming noise is extracted and compared with that of the training noises to determine the best pre-trained filter. Since this method uses power spectral density (PSD) estimates \cite{ref21} as feature representations, the filter selection accuracy degrades when the incoming noise exhibits a non-flat spectrum. In order to enhance the reliability of filter selection, several studies \cite{ref22,ref23,ref24} introduced deep learning-based methods that leverage convolutional neural networks (CNNs) to replace the PSD-based similarity matching. Furthermore, other works \cite{ref25,ref26} explored the CNN-based generative fixed-filter methods to address the limitation of having only a finite number of pre-trained filters in the SFANC approach. Nevertheless, the implementation of deep learning-based ANC requires a co-processor to perform extensive computations on user terminals such as smartphones or laptops, which increases power consumption and limits its practicality in portable devices \cite{ref27}. 

To address the aforementioned problems, this paper proposes a Subband Architecture for Selective Fixed-filter Active Noise Control (SA-SFANC). Our key contributions are as follows: (i) a set of optimal control filters is pre-trained for each subband and stored in a designated database, with their lengths reduced via decimation \cite{ref9} to lower storage requirements; (ii) the measured signal is decomposed into subbands by an analysis filter bank, and the most suitable control filter for each subband is selected by employing Jaccard similarity \cite{ref28}; and (iii) the selected sub-filters are synthesized into a fullband filter using the FFT-1 weight stacking method \cite{ref11}.

The rest of this paper is organized as follows. Section \hyperref[sec2]{II} presents a statistical analysis of the fixed-filter ANC system. Section \hyperref[sec3]{III} provides a detailed description of the SA-SFANC approach. In Section \hyperref[sec4]{IV}, experimental results are presented to evaluate the performance of our proposal. Finally, conclusions are given in Section \hyperref[sec5]{V}.

\section{Statistical Analysis for Fixed-Filter ANC}
\label{sec2}
Fixed-filter methods involve two main phases: a pre-training stage and an on-line control stage \cite{ref17,ref18,ref19,ref20}. For the former, a real-valued training noise $x_\text{t}(n)$ is first fed
into the system as a reference signal. Its frequency-domain representation is given by \cite{ref17}
\begin{equation}
\label{eq1}
    X_\text{t}(\omega) = T_\text{t}(\omega) \cdot \text{rect}\left(\frac{\omega-\omega_\text{t}}{2B_\text{t}} \right),
\end{equation}
where $T_\text{t}(\omega)$ represents the spectral shaping function of $X_\text{t}(\omega)$. The rectangular function is defined as
\begin{equation}
\label{eq2}
    \text{rect}\left(\frac{\omega-\omega_\text{t}}{2B_\text{t}} \right) = \begin{cases} 
1 & \left| \omega \right| \in [\omega_\text{t} - B_\text{t}, \omega_\text{t} + B_\text{t}] \\ 
0 & \text{otherwise},
\end{cases}
\end{equation}
where $\omega_\text{t}$ and $2B_\text{t}$ denote the center frequency and bandwidth of the training noise, respectively. According to \cite{ref17}, once the adaptive algorithm converges, the optimal control filter for $x_\text{t}(n)$, denoted by $W_\text{t}^{\text{opt}}(\omega)$, is expressed as
\begin{equation}
\label{eq3}
    W_\text{t}^{\text{opt}}(\omega) = \frac{P(\omega)}{S(\omega)}\cdot \text{rect}\left(\frac{\omega-\omega_\text{t}}{2B_\text{t}} \right),
\end{equation}
where $P(\omega)$ and $S(\omega)$ represent the frequency responses of the primary and secondary paths, respectively. 

\begin{figure}[!t]
\centering
\includegraphics[width=2.6in]{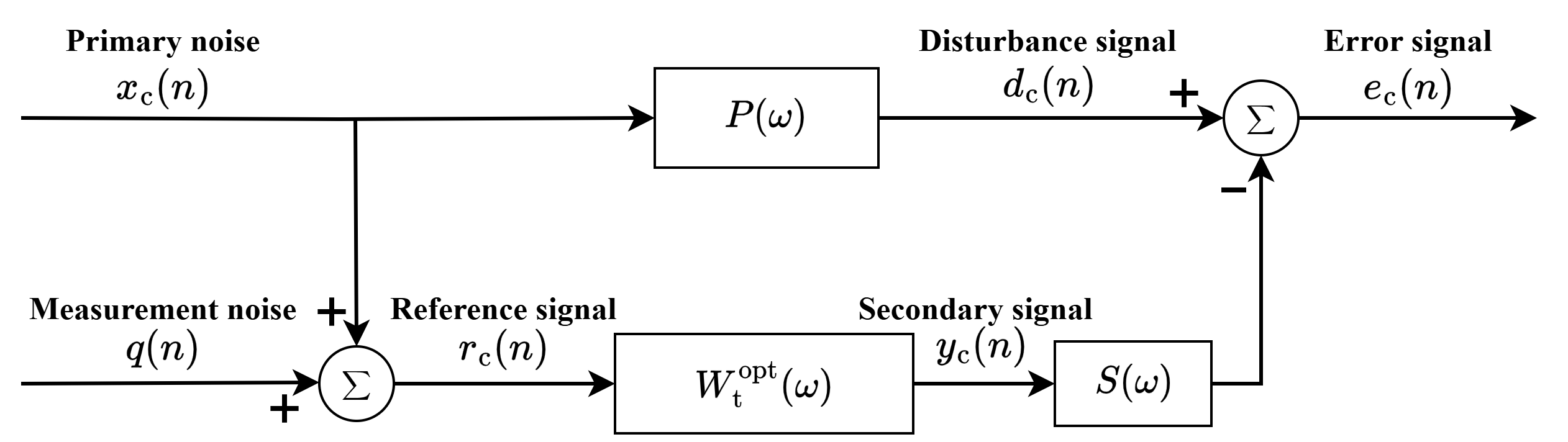}
\caption{Block diagram of the single-channel fixed-filter ANC system.}
\label{fig1}
\end{figure}

During the on-line control stage, an incoming primary noise $x_\text{c}(n)$, centered at $\omega_\text{c}$ and with bandwidth $2B_\text{c}$, is represented in the frequency domain as  
\begin{equation}
\label{eq4}
    X_\text{c}(\omega) = T_\text{c}(\omega) \cdot \text{rect}\left(\frac{\omega-\omega_\text{c}}{2B_\text{c}} \right).
\end{equation}
The optimal filter obtained in the pre-training stage is directly applied to cancel this primary noise. To illustrate this more clearly, Fig.~\ref{fig1} shows a practical scenario of a fixed-filter ANC system, where the observed reference signal is contaminated by a measurement noise from the reference microphone \cite{ref26}. The reference signal is modeled as $r_\text{c}(n)= x_\text{c}(n)+q(n)$, where $q(n)\sim \mathcal{N}(0,\sigma_q^2)$ is assumed to be a zero-mean white Gaussian process with variance $\sigma_q^2$. Accordingly, the residual error signal is computed as
\begin{equation}
\label{eq5}
    e_\text{c}(n) = d_\text{c}(n) - s(n)\circledast w_\text{t}^{\text{opt}}(n)\circledast r_\text{c}(n),
\end{equation}
where $\circledast$ denotes the linear convolution, $s(n)$ is the impulse response of the secondary path, and $w_\text{t}^{\text{opt}}(n)$ represents the impulse response of the optimal filter for $x_\text{t}(n)$. The disturbance signal is given by $d_\text{c}(n) = p(n)\circledast x_\text{c}(n)$, where $p(n)$ is the impulse response of the primary path. Without loss of generality, assuming that $\omega_\text{t}=\omega_\text{c}$, $B_\text{t} \ge B_\text{c}$ and $P(\omega)$ has unit gain, the mean square error (MSE) of $e_\text{c}(n)$ can be derived as follows \cite{ref17}:
\begin{equation}
\label{eq6}
    \mathbb{E}\{e_\text{c}^2(n)\} = \frac{1}{2\pi}\int^\pi_{-\pi}S^{\text{min}}_{e_\text{c}}(\omega)d\omega+\frac{\sigma_q^2}{\pi}(B_\text{t}-B_\text{c}),
\end{equation}
where $\mathbb{E}\{\cdot\}$ denotes the expectation operator and $S^{\text{min}}_{e_\text{c}}(\omega)$ is the PSD of the error signal when the optimal filter for $x_\text{c}(n)$ is applied. It is observed from (\ref{eq6}) that when $B_\text{t} = B_\text{c}$, the MSE is equal to the minimal mean square error (MMSE) given by the Wiener-Hopf solution \cite{ref7,ref8,ref17}. 

In the SFANC method \cite{ref17}, the database stores a collection of optimal filters, each trained on a specific type of noise. To select the best filter, the frequency band of the primary noise is compared with those of the training noises. However, real-world noise consists of a mixture of multiple sources, leading to performance degradation of the SFANC approach \cite{ref18}.

\begin{figure}[!t]
\centering
\includegraphics[width=2.4in]{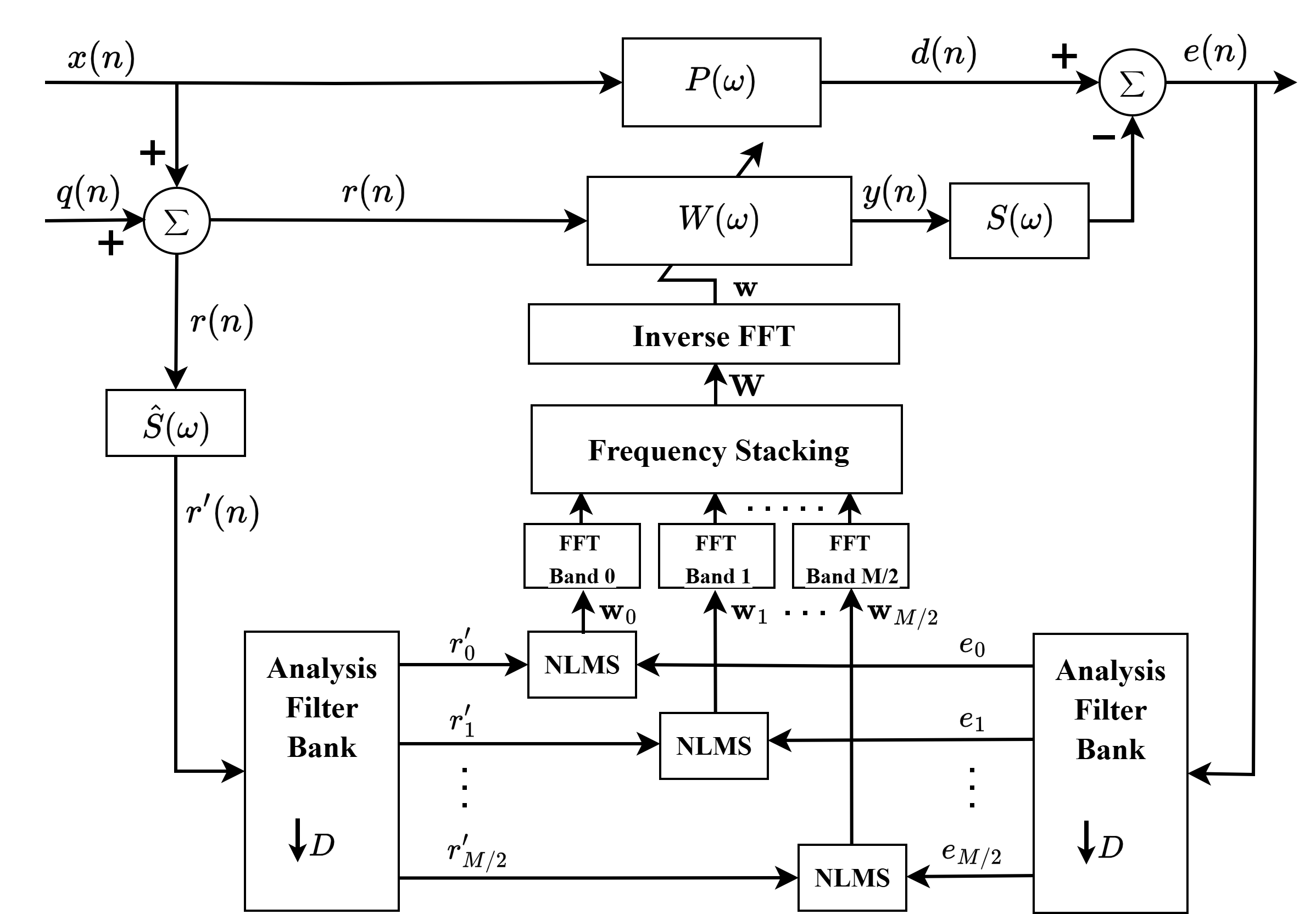}
\caption{Block diagram of the single-channel subband FxNLMS algorithm.}
\label{fig2}
\end{figure}

\section{Subband Architecture Assisted SFANC}
\label{sec3}
Fig.~\ref{fig2} illustrates the general structure of the subband adaptive filtering for FxNLMS (SAF-FxNLMS) \cite{ref11,ref12,ref13}. The secondary signal is calculated as
\begin{equation}
\label{eq7}
    y(n)=\mathbf{w}^\mathrm{T}(n)\mathbf{r}(n), 
\end{equation}
where $\mathbf{w}(n)=[w_0(n), w_1(n), \ldots, w_{L-1}(n)]^\mathrm{T} \in \mathbb{R}^{L}$ is the coefficient vector of the fullband control filter, $\mathbf{r}(n)=[r(n), r(n-1), \ldots, r(n-L+1)]^\mathrm{T} \in \mathbb{R}^{L}$ contains the most recent $L$ samples of the reference signal, and the superscript $\mathrm{T}$ denotes the transpose operation. The residual error is given by $e(n)=d(n)-s(n)\circledast y(n)$, and the filtered reference signal is computed as $r'(n) = \hat{s}(n)\circledast {r}(n)$, where $\hat{s}(n)$ is an estimate of $s(n)$. Utilizing the polyphase FFT technique \cite{ref11}, an analysis filter bank comprising $M$ single-sideband band-pass filters can be efficiently constructed from a low-pass prototype filter. Hence, the decimated output of the $m$th analysis filter is expressed as \cite{ref11} 
\begin{equation}
\label{eq8}
    r'_m(nD) = \sum^{K-1}_{k=0}r'(nD-k)a_ke^{j2\pi km/M}, m=0,1,\ldots,M/2,
\end{equation}
where $a_k$ is the $k$th coefficient of the prototype filter, $K$ is the length of the prototype and analysis filters (an integer multiple of $M$), and $D=M/2$ is the decimation factor. Note that only the first $M/2+1$ subbands are employed due to the conjugate symmetry in real-valued signals. The decimated error signal in the $m$th subband, denoted by $e_m(nD)$, is computed in a similar manner. Then, the complex NLMS algorithm updates the sub-filters according to 
\begin{equation}
\label{eq9}
    \mathbf{w}_m(n+D) = \mathbf{w}_m(n) + \mu \frac{\mathbf{r'}_m^*(n) e_m(n)}{\mathbf{r'}_m^\mathrm{H}(n)\mathbf{r'}_m(n)+\epsilon},
\end{equation}
where $\mathbf{w}_m(n)$ is the coefficient vector of the $m$th subband control filter with length $L_{\text{s}}=L/D$, and $\mathbf{r}_m'(n)=[r'_m(n),r'_m(n-D),\ldots,r'_m(n-L+D)]^\mathrm{T} \in \mathbb{C}^{L_{\text{s}}}$ contains the most recent $L_{\text{s}}$ samples of the filtered subband reference signal. Here, $\mu$ is the step size, $\epsilon$ is a small positive constant to prevent division by zero, and the superscripts $*$ and $\mathrm{H}$ denote the complex conjugate and Hermitian transpose, respectively. To reconstruct the fullband control filter, the FFT-1 stacking method is typically applied \cite{ref11}. In this way, all subband weights are first transformed using an $L_{\text{s}}$-point FFT. The frequency-domain representation of the fullband control filter is obtained by 
\begin{equation}
\label{eq10}
W[l] = 
\begin{cases}
W_{\text{round}( lM/L )}[ ((l))_{2L/M}], & l \in [0, L/2) \\
0, & l = L/2 \\
W^*[L - l], & l \in (L/2, L-1]
\end{cases}
\end{equation}
where $W[l]$ is the $l$th frequency bin of the fullband control filter, $W_m[f]$ denotes the $f$th frequency bin of the $m$th subband control filter, $\text{round}(\cdot)$ indicates rounding towards the nearest integer, and $((\cdot))_\alpha$ is the modulo-$\alpha$ operator. Finally, the time-domain fullband control filter $\mathbf{w} \in \mathbb{R}^L$ is obtained by applying the $L$-point inverse FFT (IFFT) to the vector $\mathbf{W}=\left[W[0], W[1], \ldots, W[L-1]\right]^\mathrm{T} \in \mathbb{C}^L$.

\begin{figure}[!t]
\centering
\includegraphics[width=2.5in]{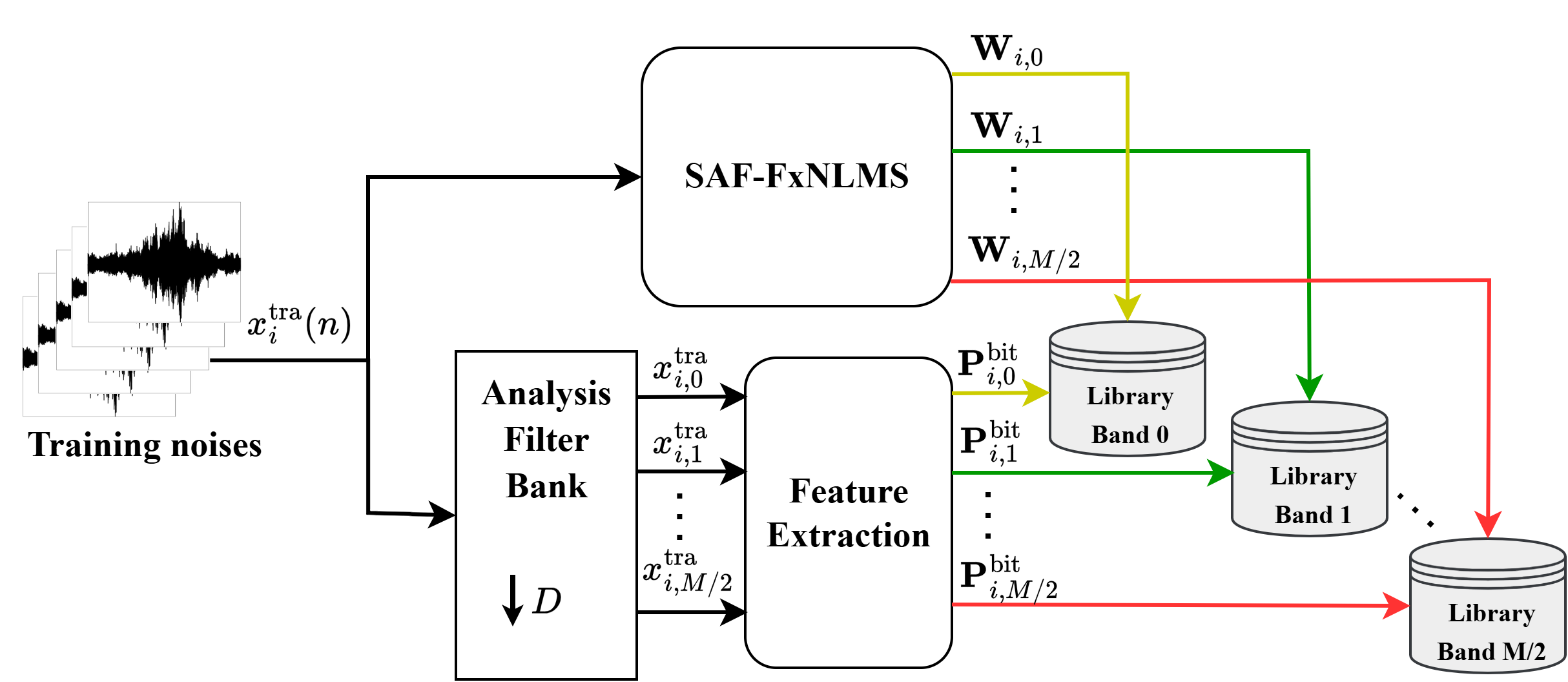}
\caption{Off-line training stage of the subband architecture for selective fixed-filter active noise control.}
\label{fig3}
\end{figure}

\begin{figure}[!t]
\centering
\includegraphics[width=2.8in]{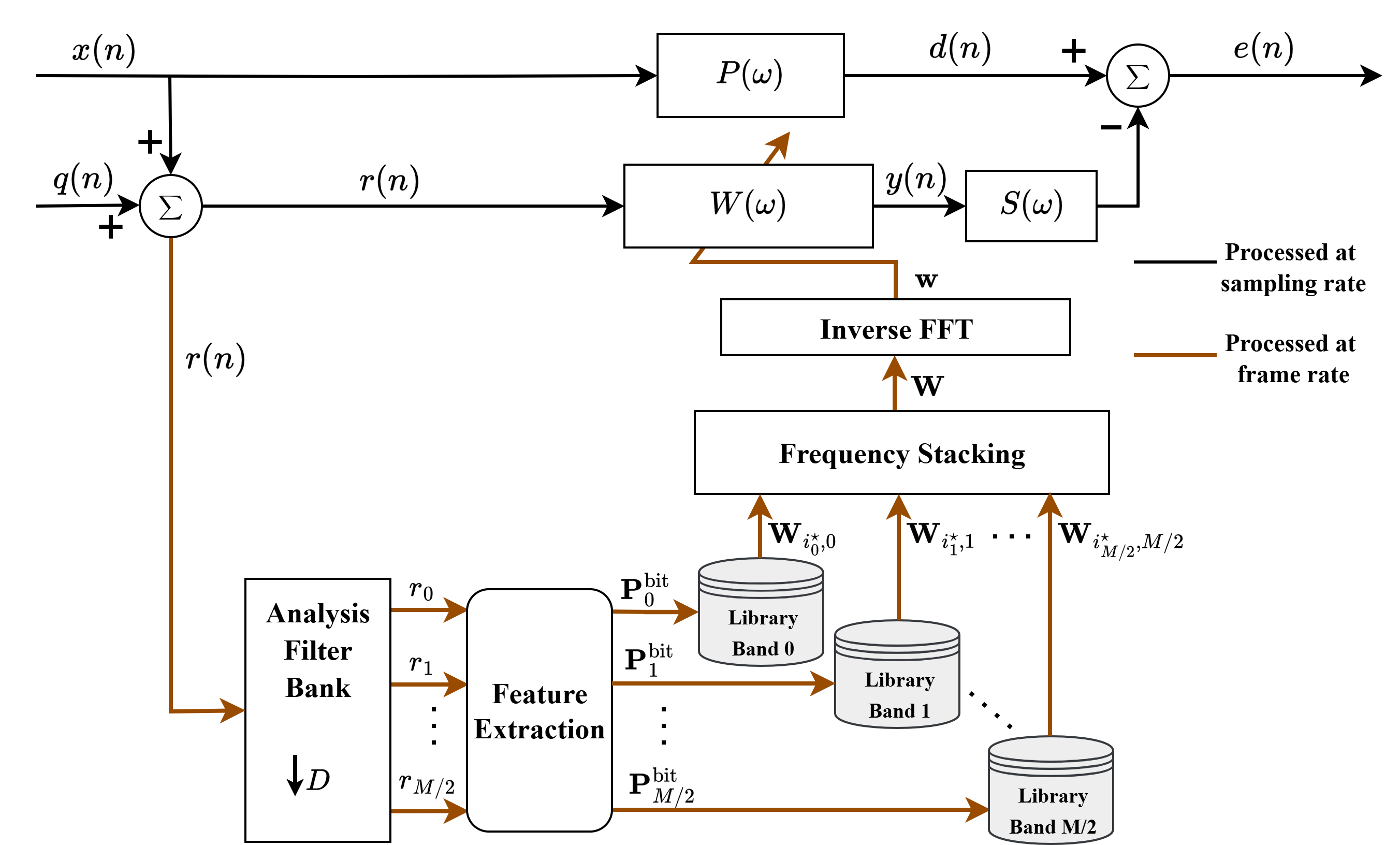}
\caption{On-line control stage of the subband architecture for selective fixed-filter active noise control.}
\label{fig4}
\end{figure}

Fig.~\ref{fig3} depicts the off-line training stage of the proposed SA-SFANC method. A collection of broadband training noises, denoted by $\{x^\text{tra}_i(n),i=0,1,\ldots,I-1\}$, is input to the ANC system as the reference signals. For each training signal $x^{\text{tra}}_i(n)$, the SAF-FxNLMS algorithm is applied. After convergence, the frequency-domain representations of the sub-filters are stored, where $\mathbf{W}_{i,m} \in \mathbb{C}^{L_{\text{s}}}$ denotes the frequency-domain weight vector of the $m$th subband filter corresponding to the $i$th training noise. In parallel, $x^{\text{tra}}_i(n)$ is decomposed using (\ref{eq8}) to obtain $\{x^{\text{tra}}_{i,m},m=0,1,\ldots,M/2\}$. The frequency-band features of each subband signal are extracted by computing the PSD estimate using Welch's method \cite{ref21}, which partitions $x^{\text{tra}}_{i,m}$ into overlapping segments. Assuming that a $V$-point FFT is performed on each segment, the resulting PSD estimate is a vector of length $V$ that retains all FFT bins. Taking the logarithm of these PSD values yields the $\log$-PSD vector for the $i$th training signal in the $m$th subband, denoted as $\mathbf{P}_{i,m} \in \mathbb{R}^V$. Since only the frequency-band occupancy information is required, $\mathbf{P}_{i,m}$ is binarized using a thresholding scheme:
\begin{equation}
\label{eq11}
[\mathbf{P}^{\text{bit}}_{i,m}]_v=
\begin{cases}
0, & [\mathbf{P}_{i,m}]_v < \mathcal{T} \\
1, & [\mathbf{P}_{i,m}]_v \ge \mathcal{T}
\end{cases}
\end{equation}
where $[\mathbf{P}^{\text{bit}}_{i,m}]_v$ denotes the $v$th element of the binary vector $\mathbf{P}^{\text{bit}}_{i,m}$ for $v \in \{0,1,\ldots, V-1\}$, and $\mathcal{T}$ is the threshold, computed as the average of the maximum and minimum values in $\mathbf{P}_{i,m}$:
\begin{equation}
\label{eq12}
\mathcal{T}=\frac{\max(\mathbf{P}_{i,m})+\min(\mathbf{P}_{i,m})}{2}.
\end{equation}
Finally, the frequency-domain weight vector and the corresponding bit vector are stored in a dedicated library (database) for each subband.

In the subsequent control stage shown in Fig.~\ref{fig4}, we consider the practical scenario, where the measurement noise $q(n)$ is included. Compared to traditional adaptive algorithms, the proposed scheme eliminates the need for secondary path estimation, error signal acquisition, and FFT computations for subband control filters. Moreover, the coefficients are updated at the frame rate (i.e., once per second) \cite{ref23}, significantly enhancing computational efficiency. Specifically, upon receiving one second of the reference signal $r(n)$, our system performs subband decomposition. Then, binary feature representations $\{\mathbf{P}^{\text{bit}}_m,  m=0,1,\ldots,M/2\}$ are obtained by estimating the $\log$-PSD vectors \cite{ref21} and applying binarization in (\ref{eq11}). These bit vectors are compared with those stored in the database to compute similarity scores. For each subband, the index of the most matched pre-trained filter is determined by
\begin{equation}
\label{eq13}
i^\star_m= \underset {i \in \{0, 1, \ldots, I-1\}} {\text{arg max}} J\left(\mathbf{P}^{\text{bit}}_m,\mathbf{P}^{\text{bit}}_{i,m}  \right),
\end{equation}
where $J(\mathbf{a},\mathbf{b})$ denotes the Jaccard similarity \cite{ref28} between binary vectors $\mathbf{a}$ and $\mathbf{b}$. Afterward, the selected subband control filters $\{\mathbf{W}_{i^\star_m,m},m=0,1,\ldots,M/2\}$ are stacked using (\ref{eq10}) to synthesize $\mathbf{W}$. Finally, $\mathbf{w}$ is obtained by taking the $L$-point IFFT of $\mathbf{W}$, and is used to process the reference signal at the sampling rate.

\section{Experimental Results}
\label{sec4}
\begin{figure}[!t]
    \centering
    \subfloat[]{
        \includegraphics[width=0.38\linewidth]{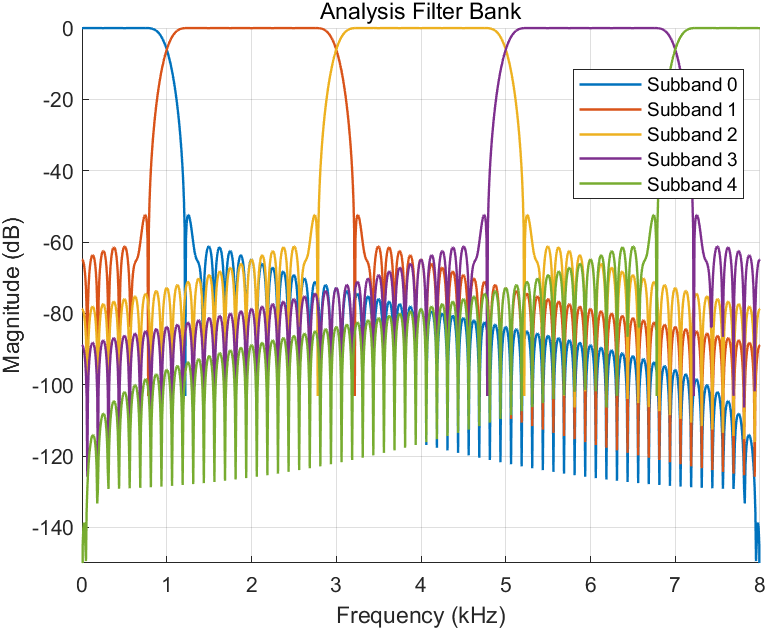}
        \label{fig5a}
    }
    \hfill
    \subfloat[]{
        \includegraphics[width=0.54\linewidth]{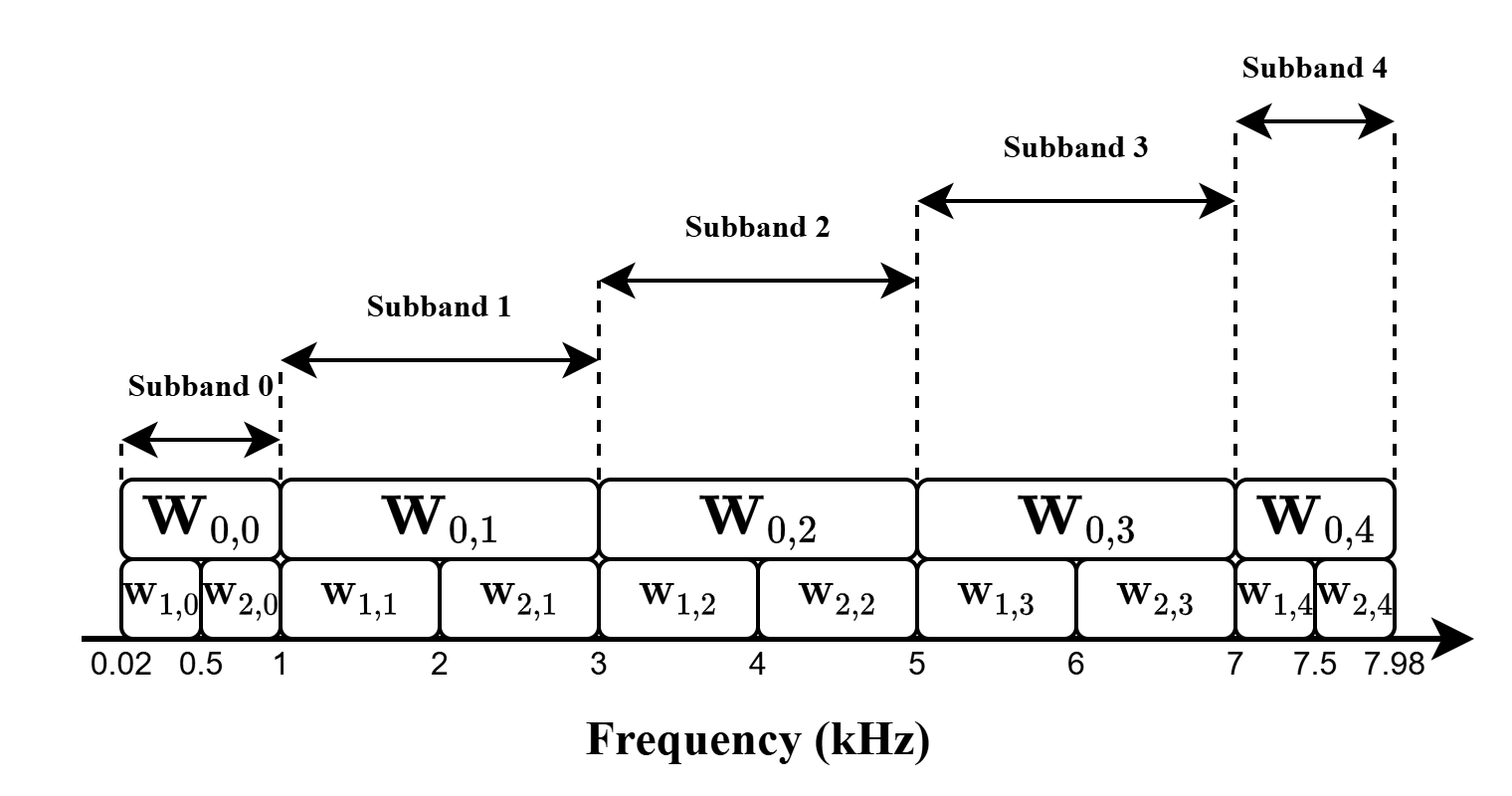}
        \label{fig5b}
    }
    \caption{(a): Frequency responses of the polyphase FFT filter bank. (b): Frequency bandwidths of the pre-trained control filters.}
    \label{fig5}
\end{figure}

Experiments were conducted to evaluate the proposed approach in comparison with the SAF-FxNLMS \cite{ref11} and SFANC \cite{ref17} algorithms. The primary and secondary paths were modeled as band-pass filters, and an ideal secondary path estimate was assumed. The fullband control filter length $L$ was 1024, and the analysis filter length $K$ was set to 128. $M=8$ was used (5 subbands were employed), and the prototype low-pass filter was designed following \cite{ref11}. At a sampling rate of 16 kHz, the frequency responses of the analysis filter bank are shown in Fig.~\ref{fig5}~\subref{fig5a}. The signal-to-noise ratio (SNR) between $x(n)$ and $q(n)$ was 20 dB. To train the sub-filters, we generated $I=3$ white Gaussian noise signals: one with a contiguous spectrum from 0.02 to 7.98 kHz, and two with multi-band spectra. This yielded 15 pre-trained control filters with different bandwidths shown in Fig.~\ref{fig5}~\subref{fig5b}. For consistency, 15 control filters were also trained for SFANC following \cite{ref22}. For feature extraction, the segment lengths for the SFANC and SA-SFANC methods were set to 256 and 64 samples, respectively. Correspondingly, 512-point and 128-point FFTs were applied to each segment \cite{ref21}. The step size $\mu$ was chosen as 0.01, and the regularization parameter $\epsilon$ was $10^{-6}$.

Under the above settings, the SAF-FxNLMS algorithm \cite{ref11} performs 4000 updates per second, whereas the SFANC method reduces it to a single update by processing at the frame rate (i.e., once per second). Similarly, our method updates once per second but has higher complexity than the SFANC method due to subband decomposition, frequency stacking, and IFFT operations. Nonetheless, the storage requirements of our proposal are reduced by subband decimation with a factor of $D=M/2=4$. Moreover, SFANC is limited to tackling only 15 noise types with its 15 pre-trained filters, while our method can synthesize $3^5 = 243$ fullband filters via weight stacking, improving adaptability to diverse noisy environments.

\begin{figure}[t]
    \centering
    \subfloat[]{%
        \includegraphics[width=0.195\textwidth]{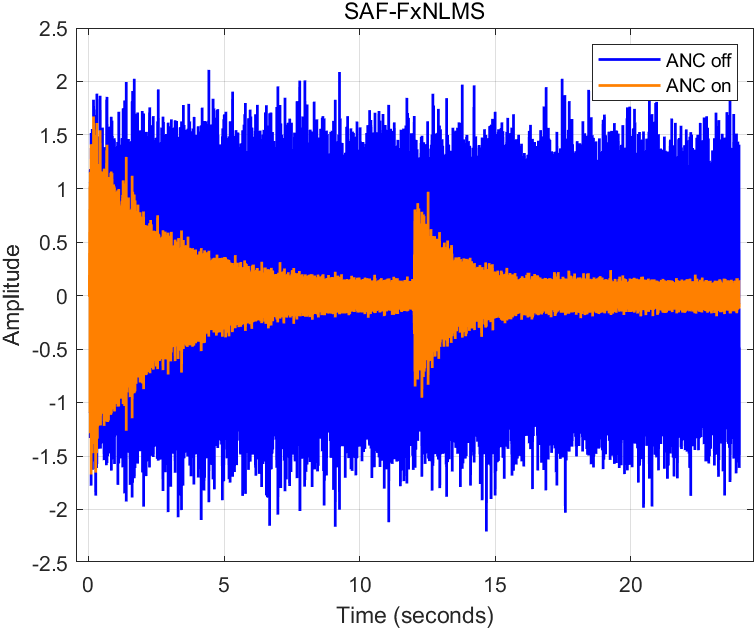}%
        \label{fig6a}
    }
    \hfill
    \subfloat[]{%
        \includegraphics[width=0.195\textwidth]{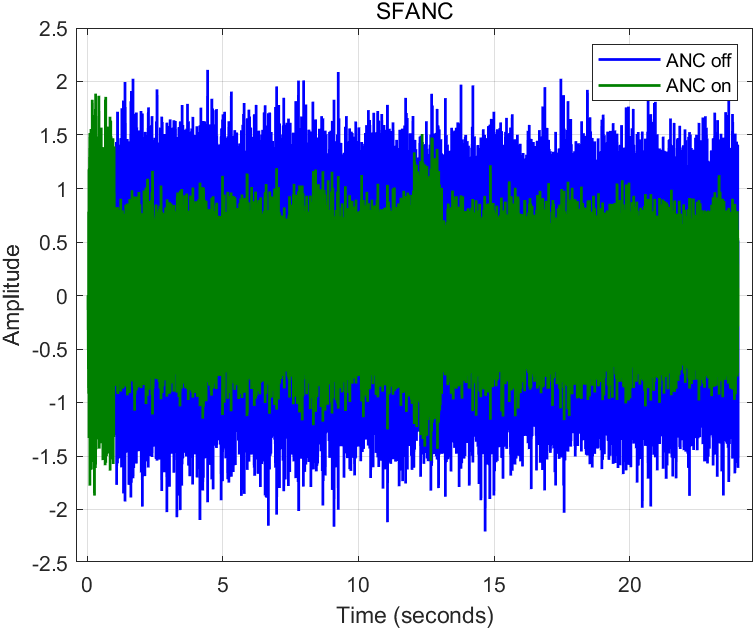}%
        \label{fig6b}
    }
    \\
    \subfloat[]{%
        \includegraphics[width=0.195\textwidth]{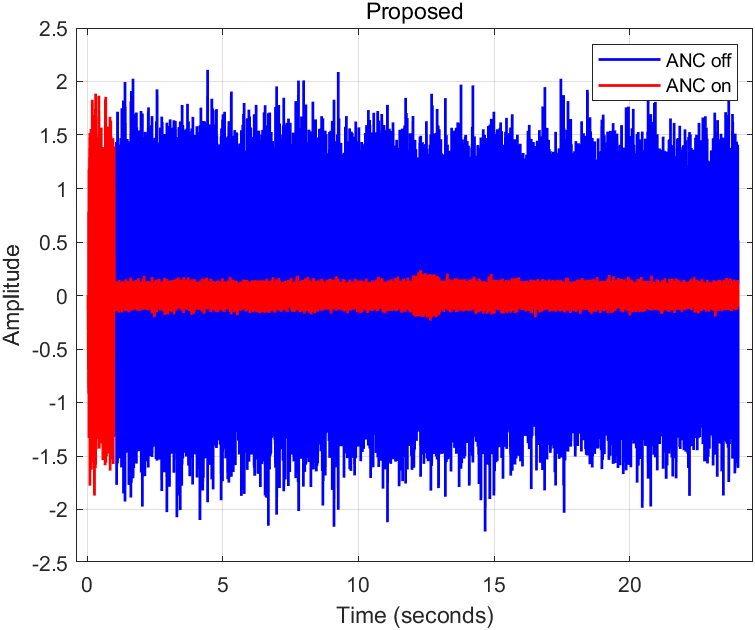}%
        \label{fig6c}
    }
    \hfill
    \subfloat[]{%
        \includegraphics[width=0.195\textwidth]{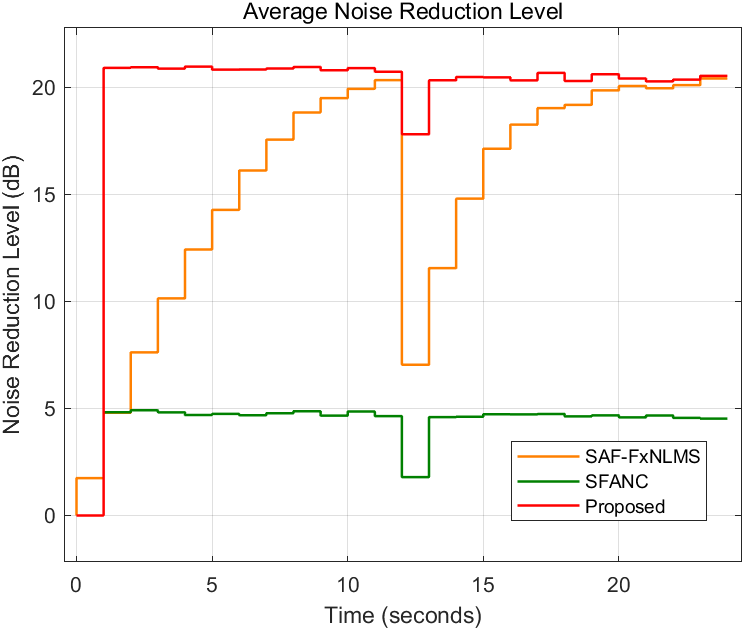}%
        \label{fig6d}
    }
    \caption{(a)-(c): Error signals of different algorithms when ANC is off and on. (d): Average noise reduction level (per second) on the simulated noise.}
    \label{fig6}
\end{figure}

Fig.~\ref{fig6} shows the results under a simulated multi-band noise scenario, where the noise power is concentrated in the 0.5--2, 3--6, and 7--7.5 kHz bands during the first 12 seconds, and then shifts to the 0.02--1, 2--5, and 6--7.98 kHz bands in the latter 12 seconds. It can be observed that the SAF-FxNLMS algorithm \cite{ref11} requires approximately 10 seconds to converge and needs to re-adapt when the primary noise changes. In contrast, the fixed-filter methods respond rapidly to such variations. However, the SFANC method \cite{ref17} is less effective than conventional adaptive algorithms due to a limited set of pre-trained filters. As shown in Fig.~\ref{fig6}~\subref{fig6d}, with the incorporation of the subband structure, a noise reduction of approximately 21 dB is achieved, and the attenuation remains robust even when the noise characteristics vary.

\begin{figure}[t]
    \centering
    \subfloat[]{%
        \includegraphics[width=0.195\textwidth]{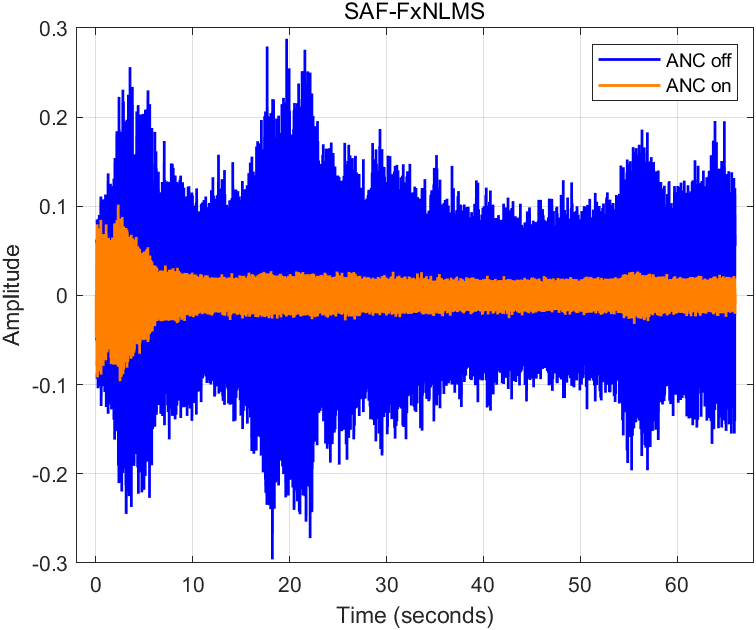}%
        \label{fig7a}
    }
    \hfill
    \subfloat[]{%
        \includegraphics[width=0.195\textwidth]{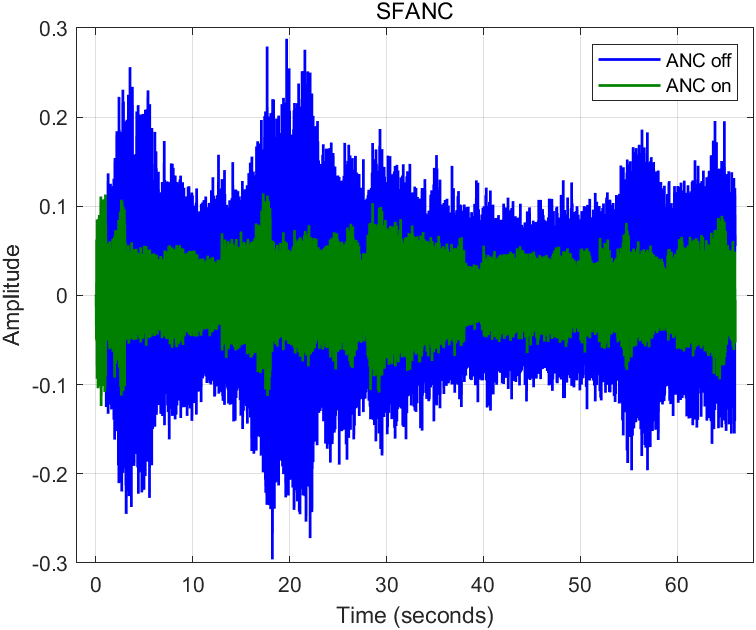}%
        \label{fig7b}
    }
    \\
    \subfloat[]{%
        \includegraphics[width=0.195\textwidth]{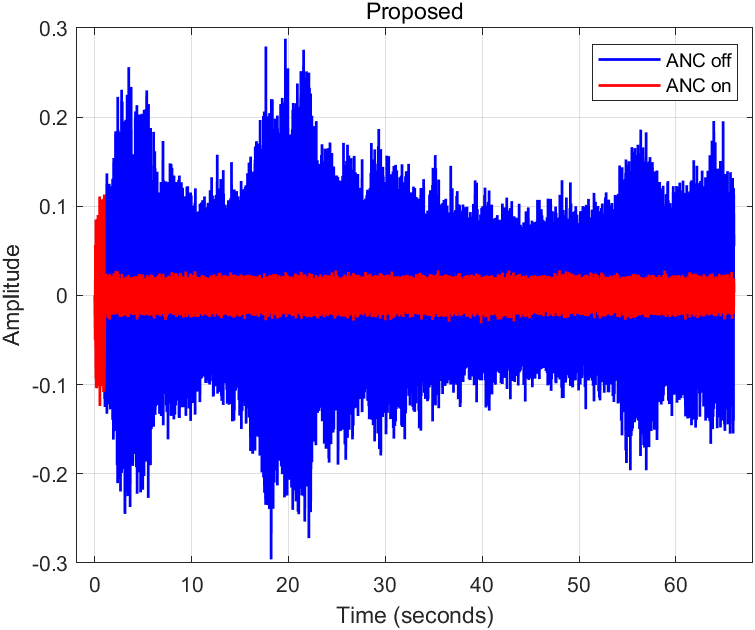}%
        \label{fig7c}
    }
    \hfill
    \subfloat[]{%
        \includegraphics[width=0.195\textwidth]{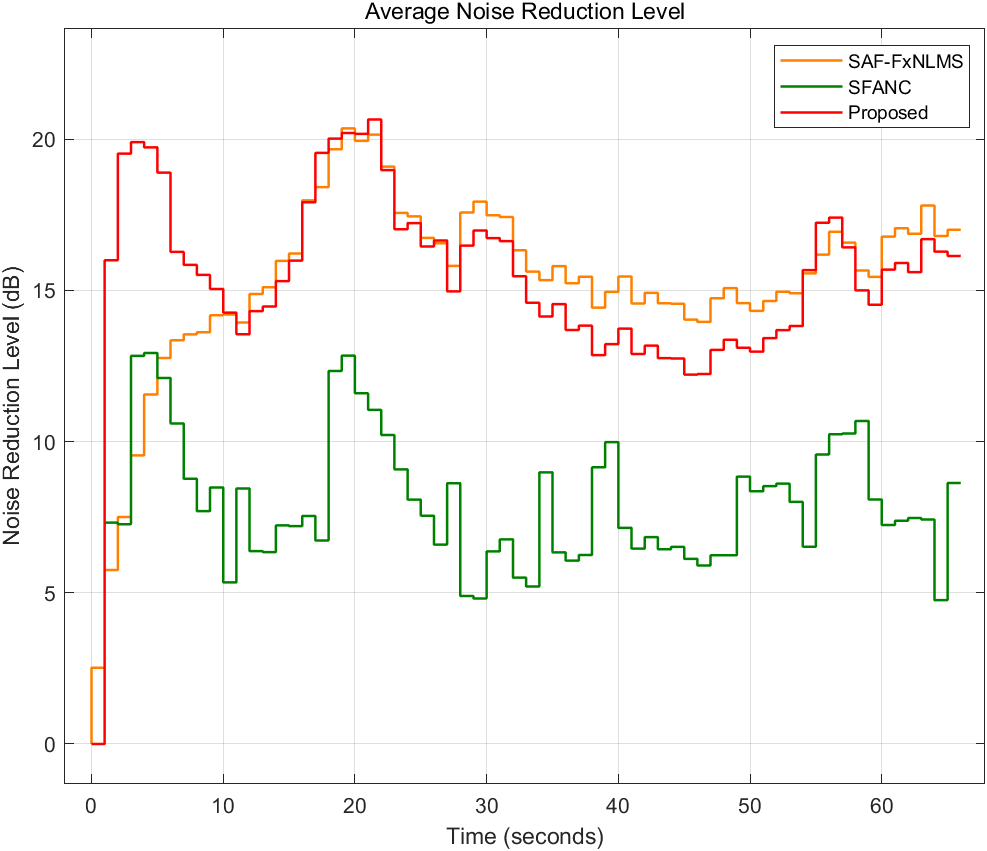}%
        \label{fig7d}
    }
    \caption{(a)-(c): Error signals of different algorithms when ANC is off and on. (d): Average noise reduction level (per second) on the real-world noise.}
    \label{fig7}
\end{figure}

Fig.~\ref{fig7} shows the performance on a real-world noise recorded in a factory environment \cite{ref29}, sourced from \textit{Freesound} \cite{ref30}. Only the first 66 seconds of this recording were used in our experiments. This factory noise exhibits a non-flat spectrum with energy primarily concentrated between 0.02 and 1.5 kHz. Unlike the original SFANC scheme \cite{ref17}, our proposal assigns the most suitable sub-filter to each frequency range of the noise, thereby consistently achieving higher noise reduction and demonstrating superior stability. The SAF-FxNLMS algorithm \cite{ref11} converges within 10 seconds and achieves the best noise suppression during the final 30 seconds. Note that increasing the number of pre-trained filters enables fixed-filter methods to handle a broader variety of noise types, but also raises computational and storage costs, highlighting a practical trade-off.

\section{Conclusions}
\label{sec5}
This paper has presented a novel fixed-filter ANC method based on a subband structure. By leveraging a pre-trained filter strategy, the proposed approach achieves improved convergence speed and noise reduction performance. Compared to the original SFANC method \cite{ref17}, our proposal significantly reduces storage requirements, while experimental results demonstrate superior robustness in complicated acoustic environments. Moreover, it shows potential for future extension, including advanced filter bank design, feature extraction, and similarity-based filter selection.

\bibliographystyle{IEEEtran}
\bibliography{references}

\end{document}